# Utilizando o Galileoscópio em observações astronômicas

Using the Galileoscope in astronomical observations

Vinicius de Abreu Oliveira*[1], Marly Aparecida da Silva[2]

[1] UNIPAMPA, Caçapava do Sul, Brasil
[2] UNIPAMPA, Caçapava do Sul, Brasil

**Resumo**

*O projeto visa atrair a atenção dos alunos e professores, do Ensino Médio Estadual no Município de Caçapava do Sul, RS, para as Ciências e, em especial, à Astronomia. Realizou-se observações astronômicas utilizando um Galileoscópio, tendo como alvo principal a Lua. Também foram observados outros objetos celestes que apresentam intenso brilho no céu noturno. A seleção de alvos e a identificação dos mesmos no céu durante as observações foi auxiliada pela utilização de um software livre de simulação de planetário, Stellarium. Os resultados obtidos foram qualitativos e de grande repercursão, pois ficou evidente o interesse dispertado nos participantes a cerca dos alvos selecionados para a observação, além de melhorar o entendimento de suas propriedades físicas diretamente observadas (e.g. cor). Por fim, este projeto evidenciou que a utilização de equipamentos simples, e de relativo baixo custo, são indispensáveis para atrair mais pessoas, em especial os jovens estudantes, para o mundo da Ciência e da Astronomia.*

*Palavras-chave: Ensino. Astronomia. Galileoscópio. Stellarium.*

**Abstract**

*This project aims to attract school students and teachers from the state education system from Caçapava do Sul - RS to Sciences and specially to Astronomy. We made astronomical observations using a Galileoscope choosing the Moon as a primary target. We also observed others objects that present hight brightness in the night sky. The selection of targets, and their identification during the observations were carried out by a free software of planetary simulation, Stellarium. The results were in qualitative form and they show the great interest demonstrated by those participating in the project. Furthermore, this project helped to improve the understanding of the physical proprieties of the night sky objects (e.g. color). Finally, the project has showed that using a simple equipment and of relatively low cost it is possible to bring more people, specially the young students, to the Science World and to Astronomy.*

*Keywords: Education. Astronomy. Galileoscope. Stellarium.*

## 1 Introdução

O estudo da astronomia é praticamente tão antigo quanto a história da humanidade. Afinal, o desejo do conhecimento e a curiosidade sempre incentivaram o desenvolvimento do Homem. Porém, antes de se formalizar como ciência, a Astronomia apresentou suas bases tanto em razões místicas, culturais quanto religiosas (SAGAN, 1996).

O início da Astronomia Moderna, i.e. Científica, normalmente é indicado como o ano de 1609, sendo o ano em que Galileu Galilei inovou nas técnicas e nos instrumentos de observação astronômicas. Sua maior contribuição foi a utilização de uma luneta para a observação astronômica. Este era um instrumento recém-desenvolvido por Hans Lippershey, um fabricante de lentes dos Países Baixos, sendo que a luneta de Galileu era composta de um tubo de chumbo e um par de lentes, uma convergente e outra divergente, apresentando um poder de aumento inicial de 3x. Após algumas adaptações, Galileu atingiu um aumento de 25x. (CLARET, 1998)

Atualmente, quatrocentos anos depois de Galileu, existe grande disponibilidade de equipamentos mais modernos do que uma luneta simples como aquela (OLIVEIRA, 2014). Porém, esta ainda desperta a curiosidade de muitos, pois possibilita uma nova visão dos objetos celestes já conhecidos por todos. Afinal, as bases místicas acerca de tais objetos ainda são percebidas quando nos deparamos com os conhecimentos não formais sobre os temas astronômicos (ALVES & ZANETIC, 2008), evidenciados pelo censo comum sobre vários fenômenos astronômicos observados.

Então, a utilização da Astronomia como fator atrativo para jovens ao mundo das ciências, está de acordo ao sugerido pelos Parâmetros Curriculares Nacionais, que enfatizam a importância das observações no ensino de Ciências. Lembrando que observar não significa apenas ver, mas perceber e encontrar detalhes nos alvos da observação (BRASIL, 1997). Sendo que estas podem ser diretas ou indiretas, mediante recursos técnicos ou seus produtos (LANGHI & NARDI, 2004). Ou seja, a utilização de um equipamento que seja simples e eficiente, serve como porta de entrada aos jovens para o mundo das descobertas. E por fim, LANGHI & NARDI (2007) citam que as observações astronômicas podem auxiliar na correção dos erros conceituais comuns nos livros didáticos, visto que estes são as primeiras, e às vezes únicas fontes de informação destes jovens, e de seus professores.

Justamente neste sentido que o trabalho atual objetivou realizar algumas atividades de observações astronômicas junto aos alunos e professores do Ensino Médio Estadual do município de Caçapava do Sul, RS. Tendo como principais alvos os corpos celestes próximos da Terra, em especial a Lua.

## 2 Materiais e métodos

Para o projeto, optou-se pela utilização de uma luneta conhecida como Galileoscópio (Fig. 01). Que consiste em uma luneta simples, muito similar àquela utilizada por Galileu, por isso o nome. Porém, ao contrário daquele utilizado em 1609, o atual apresenta a possibilidade de alterar as lentes oculares, permitindo uma ampliação variável da imagem do objeto, obtendo-se resultados melhores do que aqueles obtidos no século XVII.

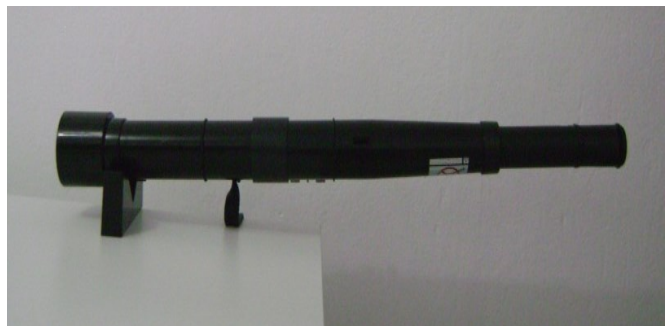

Figura 01 – Galileoscópio montado sobre uma mesa, configuração completa.

Durante o procedimento de montagem do equipamento não houve dificuldades na colocação e posicionamento das peças que constavam no kit (Fig. 02), apenas bastando seguir os passos descritos no manual que acompanha o equipamento. O kit é parte integrante do projeto Aventuras na Ciência[1], organizado em parceria pela Universidade de São Paulo (USP), Universidade Federal do Rio de Janeiro (UFRJ) e Universidade Estadual de Campinas (UNICAMP) com o patrocínio do Ministério da Educação e CAPES.

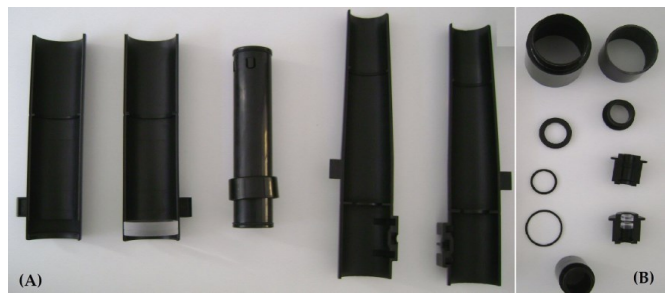

Figura 02 – Kit de montagem do Galileoscópio, (A) tubo e (B) jogo de lentes.

Para auxiliar nas observações utilizou-se um software gratuito de simulação de planetário, o Stellarium[2]. Assim, foi possível uma melhor identificação dos alvos, além de possibilitar a obtenção de informações precisas sobre os mesmos, tais como coordenadas, magnitudes aparentes e etc.

Então, o método observacional consistiu em identificar

---

[1] http://www.aventurasnaciencia.com.br/kits/astronomia
[2] http://www.stellarium.org/pt

o alvo, posicionar o Galileoscópio em direção ao objeto e realizar as observações. Aqui se tem os resultados da primeira etapa, onde se optou por realizar as observações diretamente, ou seja, sem a utilização de câmeras fotográficas ou qualquer outro equipamento de captação de informação luminosa. Esta escolha se deve apenas à vontade da equipe de aproximar ainda mais os jovens das descobertas astronômicas, em uma fase posterior pretende-se introduzir tais dispositivos e iniciar um projeto paralelo de astrofotografia.

Durante as atividades propostas, o alvo principal foi a Lua, em sua fase crescente, pois uma observação direta em fase de cheia é prejudicial à visão do astrônomo, além de dificultar a percepção de detalhes devido ao seu alto brilho. Também foram escolhidos alvos secundários para cada saída de campo, sendo os planetas do sistema solar as principais escolhas, destacando-se Vênus e Saturno. Estrelas com brilho aparente elevado, e.g. Sírius, Rígel e Betelgueuse, também foram observadas pelos participantes do projeto.

## 3 Resultados e discussões

A realização deste estudo visa, principalmente, que os participantes, após o término das atividades tenham um bom conhecimento de astronomia básica, estando em acordo com os PCN: "Identificação, mediante observações diretas, de algumas constelações, estrelas e planetas recorrentes no céu do hemisfério Sul durante o ano, compreendendo que os corpos celestes vistos no céu estão a diferentes distâncias da Terra; valorização do conhecimento historicamente acumulado, considerando o papel de novas tecnologias e o embate de ideias nos principais eventos da história da Astronomia até os dias de hoje" (BRASIL, 1998).

Sem dúvida esta meta foi alcançada, pois para as atividades propostas o equipamento se mostrou muito eficiente, em especial para observações da Lua. Sendo possível ver perfeitamente a sua superfície, com grande detalhamento das crateras e dos vales. Um resultado interessante foi a observação do planeta Vênus em fase, similarmente à Lua. Muitos dos envolvidos desconheciam esta característica dos planetas internos do Sistema Solar. Para as estrelas mais brilhantes foi possível verificar alguns detalhes fotométricos, sendo a cor (também visível a olho nu, em alguns casos) a principal característica notada. As diferenças entre magnitudes visuais também foram evidenciadas por alguns dos participantes.

Partindo destas observações foi possível explanar sobre diversas características do modelo heliocêntrico, e sobre o modelo de classificação estelar sugerido por Hiparco. A diferença de magnitude das estrelas foi associada com a distância aparente das mesmas e sua cor com a temperatura da superfície, conforme os modelos atuais. Desta forma, os participantes do projeto foram capazes de estimar distâncias aparentes relativas, ou seja, para uma determinada cor da estrela, as mais brilhantes devem estar mais próximas. Embora não seja uma análise extremamente precisa, é uma boa aproximação inicial.

Durante as etapas observacionais ocorreram pequenas dificuldades de localização espacial dos alvos, pois a pequena dimensão dos corpos celestes não contribui para identificação correta. Neste ponto, a utilização do software Stellarium solucionou muito satisfatoriamente o problema, pois facilitou a correta localização dos alvos, além de favorecer o posicionamento preciso do Galileoscópio, possibilitando que as observações fossem realizadas com sucesso. O software foi instalado em um dispositivo portátil e, utilizando a localização GPS do aparelho foi possível identificar todos os objetos do céu noturno de nosso interesse, apenas apontando o dispositivo para a região do céu. No caso deste estudo, optou-se por um tablet, devido ao maior tamanho da tela, porém é perfeitamente funcional em um smartphone, por exemplo. Também foi utilizado um planisfério celeste de cartolina como material de apoio ao tablet, que se mostrou igualmente útil e de fácil utilização.

É interessante notar que os participantes possuíam uma boa quantidade de conhecimentos oriundos da educação não formal sobre os temas abordados, e que, quando comparados com a educação formal, alguns pontos não correspondiam à realidade científica atual. Alves & Zanetic (2008) afirmam que este conhecimento prévio pode ser utilizado como base para uma melhor compreensão das informações científicas sobre astronomia, especialmente quando confrontado com experimentação. Este projeto se torna ainda mais interessante se considerarmos que alguns dos estudantes participantes só terão contato com esta educação formal de astronomia, e de ciências naturais, durante o período escolar fundamental, pois após se especializarão em outras áreas de conhecimento.

Importante ressaltar que observações diretas do Sol, sem a utilização de técnicas adequadas, são extremamente perigosas e proibitivas aos astrônomos. Logo, tais observações não foram realizadas no presente trabalho.

## 4 Conclusões

Embora tenhamos apenas realizado observações da Lua na fase crescente, pois a mesma está visível na primeira metade da noite, observações em fase minguante é igualmente proveitosa. Claro que neste caso a Lua estará visível na segunda metade da noite, o que pode ser um problema devido às possíveis faixas etárias dos estudantes participantes. Para os mais jovens, por exemplo, ainda é possível realizar as observações na fase nova, isto é, observações diurnas, que embora não apresente tantos detalhes quanto a noturna é prática e fácil de agendar nos horários de aula nas escolas.

Por fim, ficou evidente que o equipamento se comprovou como uma ferramenta simples e de baixo custo,

porém poderosa. Sob uma análise puramente qualitativa, percebeu-se que equipamentos deste tipo podem ser utilizados com sucesso, além de serem de grande importância, para a divulgação da ciência, notadamente a astronomia, podendo auxiliar na educação científica do jovem.

Para tanto, é necessário saber utilizar a ferramenta, e neste quesito este projeto se destaca pois permite a observação de diversos corpos celestes através do uso de um equipamento de fácil manuseio. Desta forma se cria uma cultura científica que pode ser difundida pela sociedade, principalmente nas escolas, pois a empolgação dos alunos durante as observações era evidente.

## Agradecimentos

## Referências